\documentclass[cleqn,10pt]{wlscirep}
\usepackage[utf8]{inputenc}
\usepackage[T1]{fontenc}
\usepackage{epsfig}
\usepackage{graphicx}
\usepackage{amsmath,amssymb}
\usepackage{bm}
\usepackage{multirow}
\usepackage{epstopdf}
\usepackage[normalem]{ulem}

\title{Tailoring interfacial effect in multilayers with Dzyaloshinskii-Moriya interaction by helium ion irradiation}

\author[1]{A.~Sud}
\author[2]{S.~Tacchi}
\author[1,3]{D.~Sagkovits}
\author[3]{C.~Barton}
\author[4]{M.~Sall}
\author[5]{L. H.~Diez}
\author[1]{E.~Stylianidis}
\author[3]{N.~Smith}
\author[3]{L.~Wright}
\author[6]{S.~Zhang}
\author[6]{X.~Zhang}
\author[4,5]{D.~Ravelosona}
\author[7]{G. Carlotti}
\author[1]{H.~Kurebayashi}
\author[3]{O.~Kazakova}
\author[1,3,*]{M.~Cubukcu}
\affil[1]{London Centre for Nanotechnology, University College London, 17-19 Gordon Street, London, WC1H 0AH, United Kingdom}
\affil[2]{Istituto Officina dei Materiali del CNR (CNR-IOM), Sede Secondaria di Perugia, c/o Dipartimento di Fisica e Geologia, Università di Perugia, I-06123 Perugia, Italy}
\affil[3]{National Physical Laboratory, Teddington, United Kingdom }
\affil[4]{Spin-Ion Technologies, Palaiseau, France}
\affil[5]{Centre de Nanosciences et de Nanotechnologies, Orsay, Île-de-France, France}
\affil[6]{King Abdullah University of Science and Technology Physical Sciences and Engineering Division, Thuwal, Makkah, Saudi Arabia}
\affil[7]{Dipartimento di Fisica e Geologia, Università di Perugia, Via Pascoli, I-06123, Perugia, Italy}

\affil[*]{murat.cubukcu@npl.co.uk}

\begin{abstract}
\textbf{We show a method to control magnetic interfacial effects in multilayers with Dzyaloshinskii-Moriya interaction (DMI) using helium (He$^{+}$) ion irradiation. We report results from SQUID magnetometry, ferromagnetic resonance as well as Brillouin light scattering results on multilayers with DMI as a function of irradiation fluence to study the effect of irradiation on the magnetic properties of the multilayers. Our results show clear evidence of the He$^{+}$ irradiation effects on the magnetic properties which is consistent with interface modification due to the effects of the He$^{+}$ irradiation. This external degree of freedom offers promising perspectives to further improve the control of magnetic skyrmions in multilayers, that could push them towards integration in future technologies.}
\end{abstract}
\begin{document}

\flushbottom
\maketitle
%
%


\section*{Introduction}

Logic devices based on nano-magnetism show the capability to couple ultrafast reversal of the magnetic state (or bit) with a non-volatile nature due to high thermal stability. A promising magnetic memory architecture is domain-wall based racetrack memory where the bit is encoded in the form of magnetic domain-walls and moved by spin currents, is still struggling in its technological implementation. One of the main reasons is related to the pinning effects introduced by defects, limiting the dynamics and thus the energy cost to store the information \cite{parkin2008magnetic, Parkin2015}. Spintronic storage or logic based on magnetic skyrmions are anticipated to be more efficient and to have a higher storage capacities \cite{roessler2006spontaneous, muhlbauer2009skyrmion, nagaosa2013topological, fert2013skyrmions, moreau2016additive, boulle2016room, woo2016observation, das2019observation, srivastava2018large, schott2017skyrmion, cubukcu2018electrical}. In fact, the small size of skyrmions allow for a significant reduction of the spacing between the bits and improves the ratio between the information flowing and the current density employed for the motion. The skyrmion topology, characterized by an integer whirling number \cite{nagaosa2013topological}, makes them particularly robust to the external environment and notably defects. Specifically, a multi-bit storage device using skyrmions as information carriers (bits), where the state of the device is modulated by an electric current that shifts the skyrmions in and out of the device. Such a behaviour, in which the state of the system (``weight") can be dynamically adapted to the environment, is analogous to a biological synapse and their synaptic plasticity \cite{huangnanotechnology}. This also opens a way for utilising such devices in novel applications, such as skyrmion-based artificial synapses and neuron type devices in low-power neuromorphic computing, or brain-inspired architectures for deep machine learning applications \cite{li2017magnetic, grollier2016spintronic, huang2018interface, zhang2016magnetic, chen2018compact, 8398530}. 

Indeed, materials lacking an inversion symmetry centre and with high spin-orbit coupling, can in general possess an additional term in the exchange energy, which is anti-symmetric; the Dyaloshinskii-Moriya interaction (DMI) \cite{moriya1960anisotropic, dzyaloshinsky1958thermodynamic}. This type of interaction favours the perpendicular alignment of adjacent spins and can lead to the formation of topological structures, such as skyrmions, possessing a particular chirality. When interfacial DMI in thin films is utilized for the formation of skyrmions, it is possible to tailor nucleation processes and skyrmion properties using a variety of approaches \cite{cubukcu2016dzyaloshinskii}. In this way, relevant magnetic properties, such as perpendicular magnetic anisotropy (PMA) or the DMI strength can be strongly modified. However, a method to reliably control the stability of skyrmions and their motion is still missing. One possible tool is the irradiation of magnetic multilayers by light irradiation using helium (He$^{+}$) ions \cite{juge2021helium}. 

In particular, it is possible to fine tune the PMA and DMI in ultrathin films and multilayers, which are dominated by interfacial effects, by appropriately adjusting the irradiation fluence \cite{balk2017simultaneous, herrera2015controlling, diez2019enhancement, devolder2013irradiation, nembach2020tuning}. Light He$^{+}$ ions, with energy of approximately 10 keV, are able to penetrate the stack, disturbing the atomic arrangement in their path and causing atoms to be slightly displaced, as He$^{+}$ ions end up deep inside the substrate \cite{rettner2002characterization}. This series of short-range interactions leads to a mild alteration of the magnetic properties of the material, without suffering from damage that would be caused by more aggressive techniques such as surface sputtering \cite{balk2017simultaneous} or cascade collisions. This method has been shown to modify the magnetic properties of ultrathin trilayers through soft intermixing \cite{chappert1998planar, herrera2015controlling, zhao2019enhancing}. Masked irradiation \cite{chappert1998planar} or focused ion beams \cite{Fassbender2008} are techniques that can be utilised to generate spatial modulation of magnetic properties in ultrathin films. He$^{+}$ ion irradiation can thus be employed as a method to pattern the magnetic properties without physical etching that would introduce defects at boundary edges of the nanostructures. Such techniques have been used to nucleate isolated skyrmions in dot geometries \cite{Fallon2020,Sapozhnikov2020} and reduce pinning centers via interfacial smoothing \cite{zhao2019enhancing,Devolder2001, Cayssol2005}.

Here, we present a way of controlling magnetic interfacial effects in multilayers exhibing DMI using He$^{+}$ ion irradiation. We report SQUID magnetometry, ferromagnetic resonance (FMR) and Brillouin light scattering (BLS) results on a set of Ta(4.5nm)/[Pt(4.5nm)/Co(1.2nm)/Ta(2.5nm)]$_{20}$ multilayers as a function of He$^{+}$ irradiation fluence (IR), ranging from $2\times 10^{\rm 14} cm ^{\rm -2}$ to $20\times 10^{\rm 14} cm ^{\rm -2}$. We study the effect of the irradiation on the magnetic properties of multilayers such as magnetic anisotropy, Gilbert damping and DMI strength. Our results show clear evidence of the effect of He$^{+}$ irradiation on the magnetic properties which is consistent with a controllable interface modification. The ability to precisely tailor the magnetic properties by externally adjusting the properties of thin-films offers a promising way to control magnetic skyrmions, making them a viable option for future technological applications.

\section*{Results}

\subsection*{SQUID magnetometry measurements}

In order to investigate the magnetic properties such as the saturation magnetization $M_{\rm s}$ of the multilayers, we performed SQUID magnetometry at room temperature, applying an external field along the in-plane and out-of-plane directions with respect to the sample surface. In Figures 1(a) and 1(b), we show the magnetization $M$ as a function of the magnetic field $\mu_{\rm 0}H$ in both orientations for different IR, which indicates that the easy axis lies along the normal to the sample. In addition, we found that the saturation magnetization maintains a constant value $M_{\rm s}$= 700 $\pm$20 kA/m within the experimental error as a function of IR.

\begin{figure}
\centering
\includegraphics[width=0.75\textwidth]{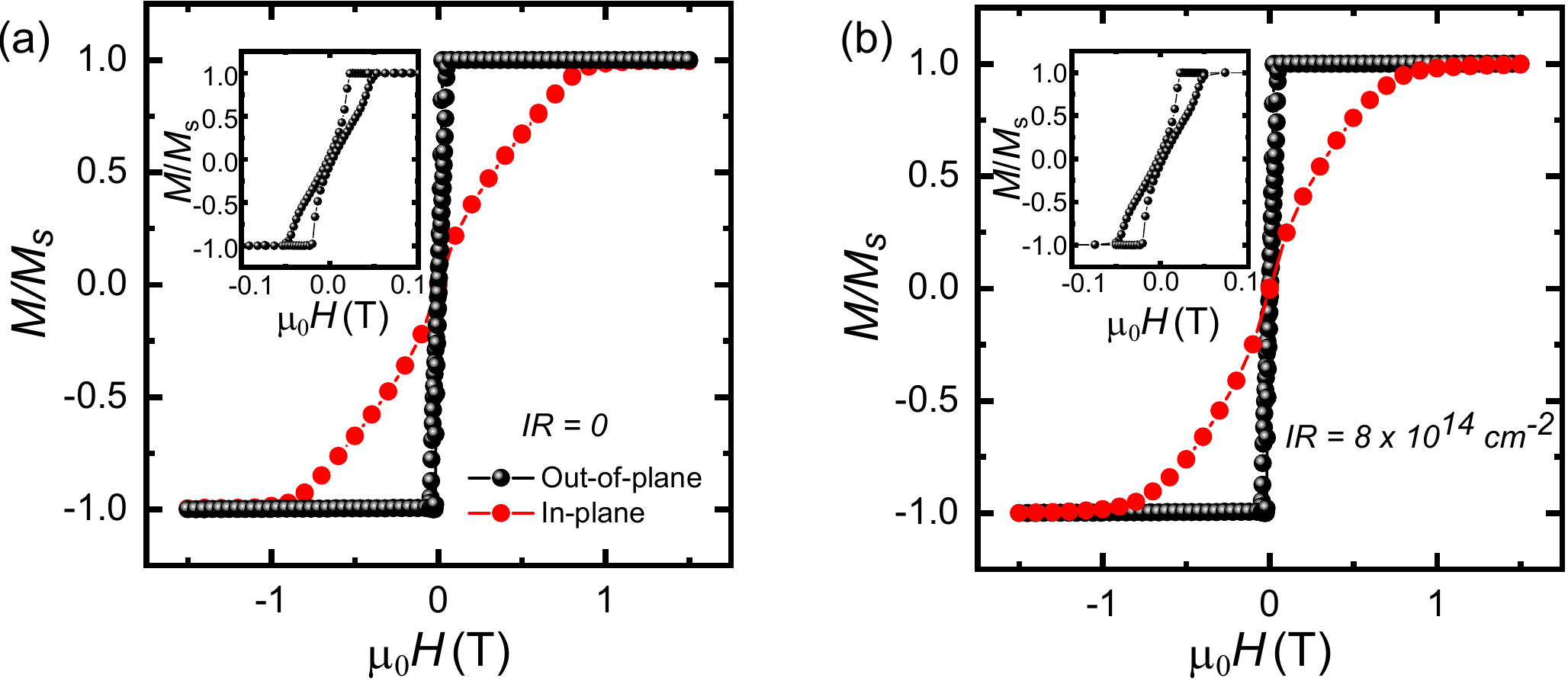}
\caption{\label{fig:one} (a-b) Normalized magnetization measurements $M$/$M_{s}$ while sweeping the external magnetic field $\mu_{\rm 0}H$ in the direction perpendicular (out-of-plane) and parallel (in-plane) to the multilayers with different IR at room temperature. Inset: Low field region for the out-of plane direction. }
\end{figure}.

\subsection*{Ferromagnetic resonance (FMR) measurements}

 In order to understand the changes to the magnetic properties due to IR, FMR was employed (Fig.2(a)) to determine the effective magnetisation $\mu_{\rm 0}M_{\rm eff}$, the effective anisotropy field $\mu_{\rm 0}H_{\rm K^{\rm eff_{\rm (FMR)}}}$ and the Gilbert damping factor $\alpha$ of the multilayers. Figure 2(b) shows the typical FMR absorption spectra for the multilayers. The absorption peaks are fitted with symmetric and anti-symmetric Lorentzian functions at each frequency respectively. The fitting formula used to extract the parameters is given in Eq.S1 (see supplementary information). By analysing these absorption peaks, the resonance position and the linewidth can be determined from which it is possible to obtain the dynamic properties for a given magnetic sample. The frequency dependent FMR results, (8 - 20 GHz), are shown in Fig.3(a-d) where the external field applied along the film normal for each IR sample. 
\begin{figure}
\centering
\includegraphics[width=0.7\textwidth]{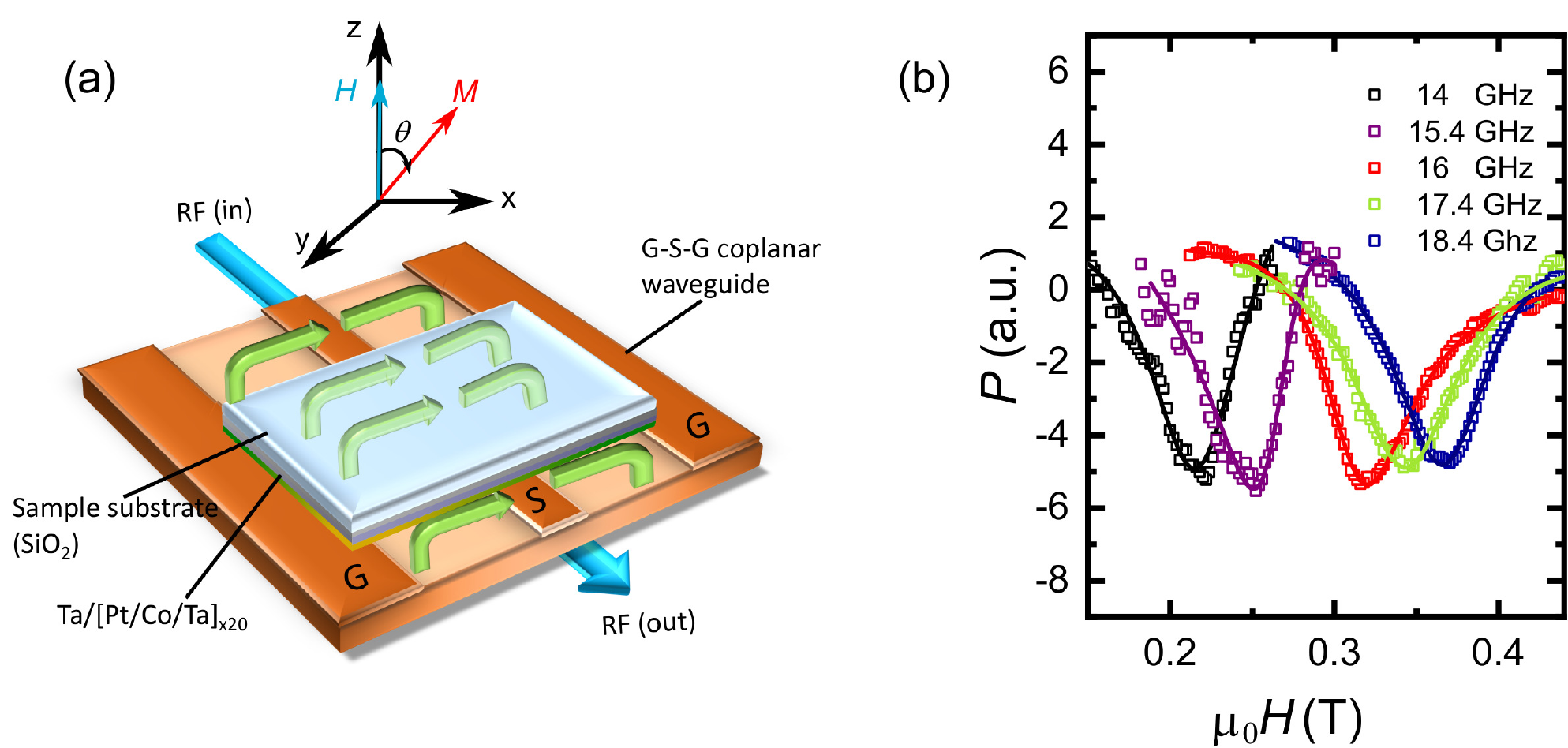}
\caption{\label{fig:two} (a) Schematic illustration of FMR setup. (b) The FMR absorption spectras for the sample IR with $2\times10^{14}cm^{-2}$ at various frequencies. }
\end{figure}
\begin{figure}
\centering
\includegraphics[width=1\textwidth]{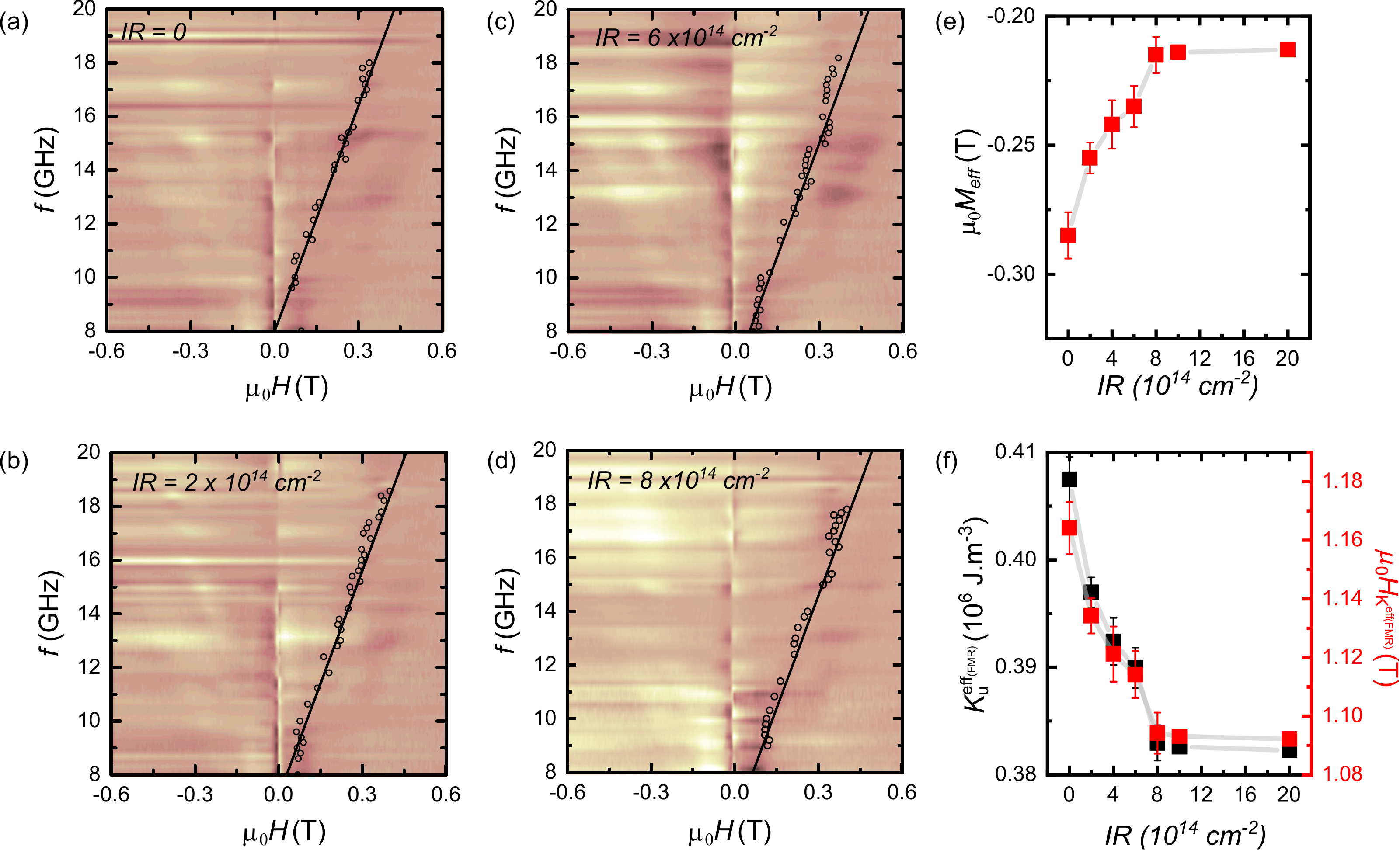}
\caption{\label{fig:three}(a-d) Microwave transmission as a function of frequency for different IR with the magnetic field perpendicular to the film plane. The black hollow markers depict the resonance field obtained by fitting FMR spectra using the Eq. S1 and solid lines are fitting curve. (e-f) The effective magnetisation $\mu_{\rm 0}M_{\rm eff}$ (e) and the effective uniaxial anisotropy field $\mu_{\rm 0}H_{\rm K^{\rm eff_{\rm (FMR)}}}$ (f) as a function of IR.}
\end{figure}

The solid lines in Fig.3(a-d), are defined by least square fitting of the resonance condition: 
$ f=\gamma\mu_{\rm 0}(H_{\rm res}- M_{\rm eff})/{2\pi}$
where $\gamma$ is the gyromagnetic ratio, $\mu_{\rm 0}M_{\rm eff} = \mu_{\rm 0}M_{\rm s}-\mu_{\rm 0}H_{\rm K^{\rm eff_{\rm (FMR)}}}$ the effective magnetization, respectively, and $\mu_{\rm 0}H_{\rm K^{\rm eff_{\rm (FMR)}}}=2K_{\rm u}^{\rm eff_{\rm (FMR)}}/M_{\rm s}$ is the effective uniaxial anisotropy field which includes first and second order magnetic anisotropy field contributions. We found that $\mu_{\rm 0}M_{\rm eff}$ is negative (Fig.3(e)), confirming that the demagnetizing energy is less than $\mu_{\rm 0}H_{\rm K^{\rm eff_{\rm (FMR)}}}$ giving rise to an easy-axis along the film normal (perpendicular easy-axis). The $\mu_{\rm 0}M_{\rm eff_{\rm (FMR)}}$ value increases compared to non-irradiated sample as shown in Fig.3(e). Fig.3(f) details the evolution of both $\mu_{\rm 0}H_{K^{\rm eff_{\rm (FMR)}}}$ and $K_{\rm u}^{\rm eff_{\rm (FMR)}}$ as a function of increasing IR. As shown $\mu_{\rm 0}H_{K^{\rm eff_{\rm (FMR)}}}$ and $K_{\rm u}^{\rm eff_{\rm (FMR)}}$ rapidly decrease upon increasing the IR up to a saturation value occurring at a IR value of $\approx$ $8\times10^{\rm 14}cm^{\rm -2}$.  We attribute this change to the interfacial modification caused by the IR. When the sample is irradiated with He$^{+}$, atomic displacements are induced which affect the related magnetic properties \cite{king2014local}. By increasing the IR, the PMA is reduced due to induced intermixing at the interface \cite{PhysRevB.62.5794,950925,doi:10.1063/1.3581896}. The Pt/Co interface roughness (or intermixing) increases linearly with IR, resulting in a continuous reduction of interfacial PMA \cite{PhysRevB.62.5794}.

\begin{figure}
\centering
\includegraphics[width=0.9\textwidth]{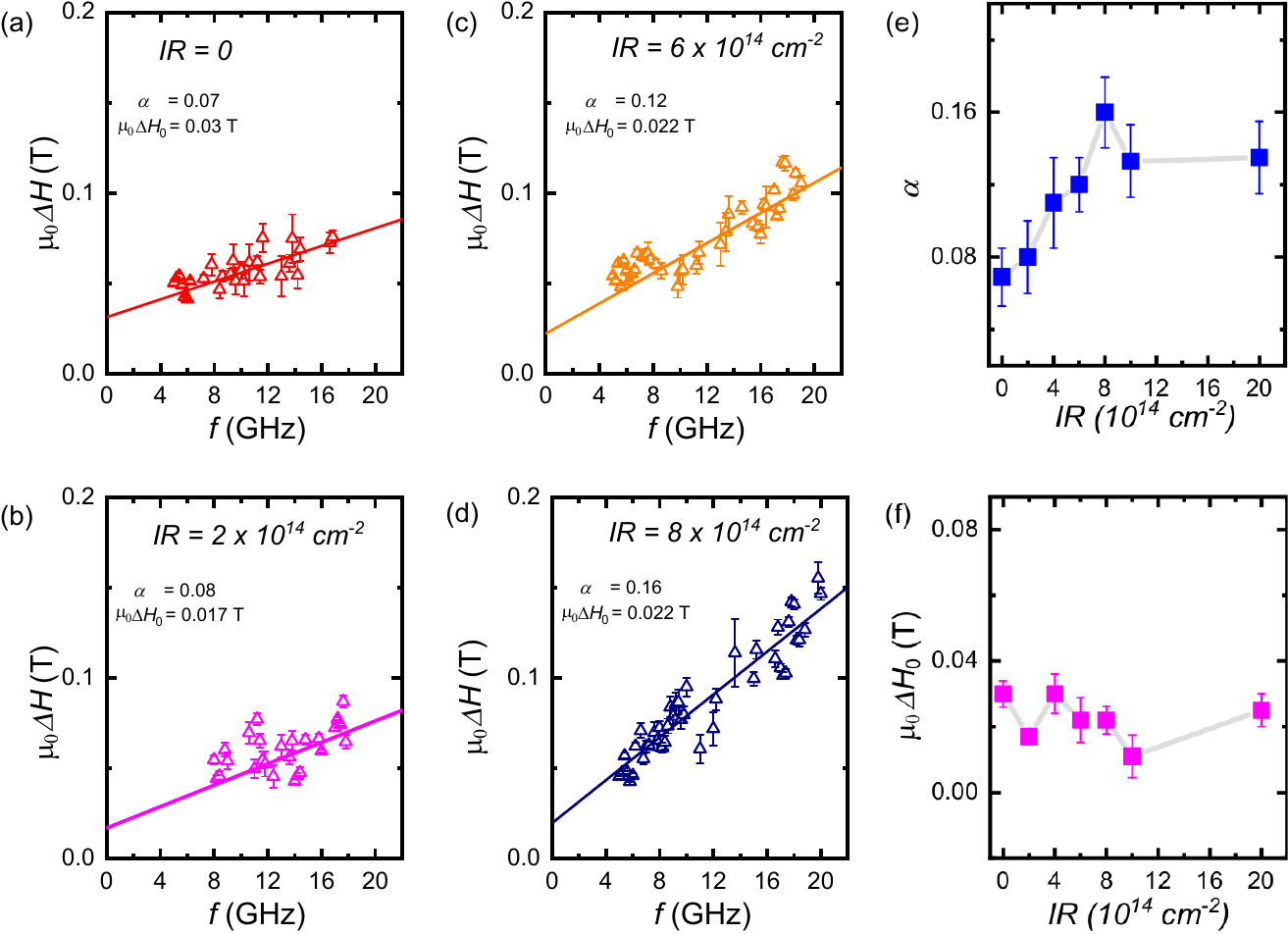}
\caption{\label{fig:four} (a-d) Dependence of the linewidth, $\mu_{\rm 0}\Delta H$, on the microwave frequency, \textsl{f}, with in the out-of-plane external magnetic field $\mu_{\rm 0}H$ for different IR. The solid lines are linear fits. (e) Plot of damping constant $\alpha$ and (f) Inhomogeneous broadening component $\mu_{\rm 0}\Delta H_{\rm 0}$ for different IR.}
\end{figure}

Analysis of the FMR linewidth $\mu_{\rm 0}\Delta H$ is a generalized method for extracting $\alpha$, which characterises the frequency dependent contribution to the Gilbert damping, as well as the inhomogeneous frequency independent contributions to the linewidth. Fig.4(a-d) shows the frequency dependence of linewidth $\mu_{\rm 0}\Delta H$ for different IR. The extrinsic contributions to linewidth: magnetic inhomogeneity and two-magnon scattering can broaden the linewidth and cause a non-linear frequency dependence. The two-magnon contribution as a cause of linewidth broadening is ruled out since the measurements were performed along the out-of-plane direction where two-magnon processes are forbidden, according to Arias-Mills theory. The data is fitted using Gilbert damping contribution and frequency independent inhomogeneous contribution using the relation, $ \mu_{\rm 0} \Delta H = \mu_{\rm 0}\Delta H_{\rm 0}+2\pi\alpha f/\gamma$
where $\alpha$ is the Gilbert damping contribution and $\mu_{\rm 0}\Delta H_{\rm 0}$ is the extrinsic contribution present due to inhomogeneities. It can be seen from Fig.4(e) that the $\alpha$ value gradually increases as the IR is increased and saturates at a certain level. This shows that the presence of radiation affects the relaxation rate, but the inhomogeneous contribution is almost constant for all the samples within the error bars (Fig.4(f)).

\subsection*{Brillouin light scattering (BLS) measurements}
\begin{figure}
\centering
\includegraphics[width=1\textwidth]{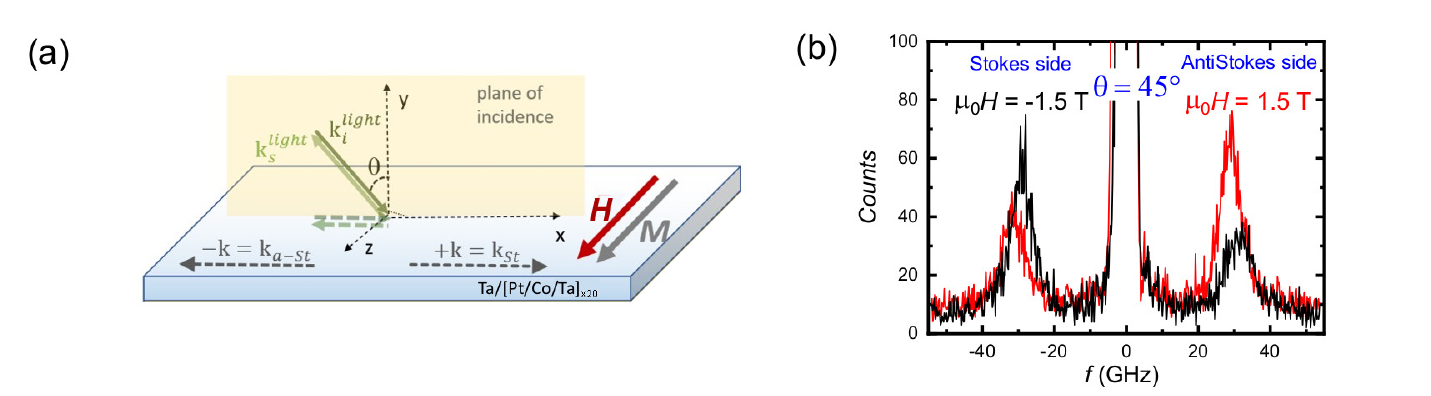}
\caption{\label{fig:five} (a) Schematic of Brillouin light scattering (BLS) experiment. The sample is saturated in-plane by an external field $\mu_{\rm 0}H$ = 1.5 T, applied along the z-axis. Stokes and anti-Stokes events in the scattering process correspond to spin waves propagating with +k and -k, respectively. (b) BLS spectra measured on the as-grown sample at an angle of incidence 45°, applying an in-plane field $\mu_{\rm 0}H$ = ±1.5 T.}
\end{figure}
As a final step in our investigation of the modifications of the magnetic properties of our multilayers induced by ion irradiation, we exploited Brillouin light scattering (BLS) with in-plane saturated samples. BLS analysis permitted us to achieve a complementary estimation of the out-of-plane anisotropy constant, and also to obtain a quantitative evaluation of the DMI strength. An in-plane magnetic ﬁeld $\mu_{\rm 0}H$ = 1.5 T, sufficiently large to saturate the magnetization in the ﬁlm plane, was applied along the z axis, while the in-plane $k$ was swept along the perpendicular direction (x-axis), corresponding to the Damon-Eshbach (DE) geometry (Fig.5(a)). Due to the conservation of momentum in the light scattering process, the magnitude of the spin wave wavevector $k$ is related to the incidence angle of light $\theta$, by the relation $k = 4\pi\sin\theta{/}\lambda $. Interfacial DMI induces a frequency asymmetry between DE modes propagating in opposite in-plane directions, corresponding to either Stokes or anti-Stokes peaks in BLS spectra (Fig.5(b)), perpendicular to the sample magnetization, following the relation $\Delta f = 2\gamma D/(\pi M_{\rm s}) k$, where $D$ is the effective DMI constant, $k$ is the spin wave vector, and $\gamma$ is the gyromagnetic ratio. In order to estimate the effective DMI constant $D$, the spin wave frequency $\Delta f$ was measured at $k$ = 1.67×10$^7$ rad/m (corresponding to $\theta = 45^\circ$) on reversing the direction of the applied magnetic field, that is equivalent to the reversal of the propagation direction of the DE spin wave mode. Fig.5(b) shows BLS spectra measured for the as-grown sample. One can observe that the Stokes and anti-Stokes peaks are characterized by a sizeable frequency asymmetry, which reverses upon reversing the direction of the applied magnetic field. 
\begin{figure}
\centering
\includegraphics[width=0.9\textwidth]{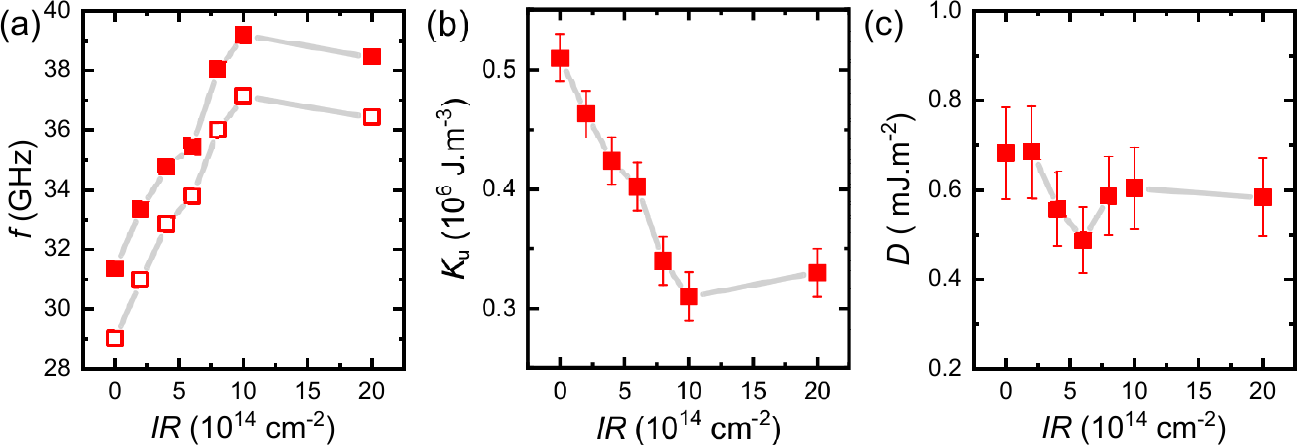}
\caption{\label{fig:six} (a) Evolution of the absolute values of the Stokes (full dots) and anti-Stokes (open dots) frequencies measured for positive applied field $\mu_{\rm 0}H$ = 1.5 T, as a function of the IR. (b) The first order out-of-plane uniaxial anisotropy constant ($K_{u}$) as a function of the IR. (c) Values of the DMI constant $D = (\pi M_{\rm{s}} \Delta f)/2\gamma k$ derived from the frequency asymmetry between the anti-Stokes and Stokes peaks $\Delta f$ of the Fig. 6(a).}
\end{figure}

In Fig.6(a) we show the evolution of the absolute values of the Stokes and anti-Stokes frequencies measured for positive applied field $\mu_{\rm 0}H$ = 1.5 T, as a function of the IR. It is evident that there is an increase of the frequencies with the IR, reflecting an increase of the effective magnetization of the stacks due to a reduction of the out-of-plane anisotropy. Fig.6(b) shows the values of the first order uniaxial anisotropy constant $K_{\rm u}$ obtained from the fit of the BLS frequency performed using the following expression that is valid for in-plane magnetized films of thickness t:
\begin{align}
 f(k) &= f_{\rm 0}(k)\pm f_{\rm DMI}(k)\notag\\
 &= \frac{\gamma \mu_{\rm 0}}{2\pi}\sqrt{H+Jk^{\rm 2}+P(kt)M_{\rm s}\left(H-\frac{2K_{\rm u}}{M_{\rm s}}+Jk^{\rm 2}+M_{\rm s}-P(kt)M_{\rm s}\right)}\pm\frac{\gamma}{\pi M_{\rm s}}Dk
\end{align}
Where $J$ is the effective exchange constant and the dipolar term $P(kt)$, in the present case of an ultrathin film, becomes $\frac{\vert kt\vert}{2}$.

\section*{Discussion}

He$^{+}$ irradiation shows a profound effect on the relaxation properties as demonstrated by a non-monotonic increase in damping value $\alpha$ (Fig.4e). The large damping parameter in these films can be attributed to an increase in surface magnon-electron interaction due to a large number of interfaces causing subsequent spin relaxation of conduction electrons that leave the Co layer at the various interfaces \cite{PhysRevB.66.214416,berger2001effect}. Furthermore, the effect of irradiation on damping in our multilayers indicates that the change in damping with IR can be correlated to a change in resistivity of the multilayers (see supplementary information, S2). The dominant magnetization relaxation in our films involves electron scattering both in the bulk and also at the interfaces. To quantify the effect of scattering as the dominant mechanism, we measure the resistivity $\rho$ of the samples (Fig.S1). We find a correlation between resistivity and damping values, both showing an increase with increasing IR. The changes in $\rho$ reflect the corresponding changes in $\alpha$. The results signify a large contribution of electron scattering on the relaxation properties of He$^{+}$ irradiated multilayers.

In agreement with the FMR results, $K_{\rm u}$ from BLS is observed to rapidly decrease on increasing IR (Fig.6(b)). One can note that for low IR values the $K_{\rm u}$ from BLS is larger than the effective uniaxial anisotropy constant obtained from FMR measurements (Fig.3(f)). This indicates that a negative contribution of the second order uniaxial anisotropy term (see supplementary information, Fig. S2), favoring an in-plane easy-axis of the magnetization, is present \cite{Mokhtari}. As can be seen in Fig.S2, the second order uniaxial anisotropy term is negative and its modulus decreases reaching an almost zero value at high IR. In Fig.6(c) we show the values of the DMI constant $D = (\pi M_{\rm{s}} \Delta f)/2\gamma k$ derived from the frequency asymmetry $\Delta f$ of the Fig.6(a). The positive value of $D$ indicates that the right-handed chirality is favored by the DMI. As IR increases, a reduction of the DMI is observed at intermediate IR  with minima at IR = $6\times 10^{\rm 14} cm ^{\rm -2}$ which then stabilises at a constant value at high IR. $Ab$ $initio$ calculations also predict that intermixing at the Pt/Co interface results in a slight diminution of the DMI \cite{Zimmermann2018}, in good agreement with our experimental results. A similar decrease with the He$^{+}$ irradiation was experimentally observed in Pt/Co/MgO \cite{juge2021helium} and W/CoFeB/MgO \cite{zhao2019enhancing} ultrathin films, although an increase was reported in Ta/CoFeB/Pt \cite{nembach2020tuning} and Ta/CoFeB/MgO \cite{diez2019enhancement,casiraghi2019bloch}. Different materials used can lead to very different interfacial structures due to robustness to intermixing, spin-orbit coupling and therefore different DMI behaviour.   

In summary, we have demonstrated the control of the interfacial properties in Ta(4.5 nm)/[Pt(4.5nm)/Co(1.2nm)/Ta(2.5nm)]$_{20}$ multilayers with DMI caused to the sample by the He$^{+}$ irradiation using SQUID magnetometry, FMR and BLS. Our results show clear evidence of i) tailoring of interface ii) different magnetic properties in multilayers with different IR. As the IR increases, we observe that the PMA decreases significantly but after the IR reaches a certain level, it approaches saturation, while a reduction of the DMI is observed at intermediate IR which then stabilises at a constant value at high IR. The He$^{+}$ irradiation induces short range atomic displacements, of the order of a few interatomic distances, leading to interface intermixing and hence altering the interface-driven PMA and DMI \cite{chappert1998planar, fassbender2004tailoring}. He$^{+}$ irradiation also shows a profound effect on the relaxation properties as demonstrated by a non-monotonic increase in damping value $\alpha$. We see a correlation between resistivity and damping values. The results signify a large contribution of electron scattering on the relaxation properties He$^{+}$ irradiated multilayers. Moreover, He$^{+}$ irradiation can be used virtually in any kind of ultra-thin materials, including ferrimagnets \cite{woo2018current, caretta2018fast, hirata2019vanishing} and synthetic antiferromagnets \cite{legrand2020room}. Finally, we should underline that we can control the magnetic properties of the thin films with DMI via He$^{+}$ ion irradiation. This has consequences for the behaviour of skyrmions. For example, their nucleation, stabilisation, size, velocity and skyrmion Hall effect. The irradiation also opens up the possibility of new writing techniques which could be used to fabricate racetracks or logic devices without physically confining skyrmions via sample edges \cite{juge2021helium}. Traditional lithographic techniques suffer from defects at boundary edges and skyrmions can be attracted to physical edges preventing efficient transport along the racetrack. This could provide a flexible way of designing countless possible structures for studying skyrmion behaviour or technological applications. The technique might be leveraged to generate a gradient of DMI, anisotropy or damping across the sample which could drive skyrmions from one side to the other with no external driving force, essentially creating a ramp for skyrmions.

\section*{Methods}

\subsection*{Sample preparation}

The multilayers Ta(4.5 nm)/[Pt(4.5nm)/Co(1.2nm)/Ta(2.5nm)]$_{20}$ were deposited simultaneously on oxidized Si substrates by DC magnetron sputtering at room temperature. They were then uniformly irradiated at room temperature using a He$^{+}$-S system from Spin-Ion Technologies with a 25 keV He$^{+}$ beam at different IR. 

\subsection*{Ferromagnetic resonance (FMR)}

The measurements were performed by placing the magnetic multilayers on a co-planar wave-guide that generates the microwave magnetic fields at various frequencies and allows for the measurement of the absorbed power in the multilayers whilst sweeping the external field. The external field was modulated with an amplitude of a few mT at 12 Hz and phase-sensitive detection was used to increase the signal to noise ratio.

\subsection*{Brillouin light scattering (BLS)}

BLS measurements were performed by focusing 150 mW of monochromatic light onto the sample surface. This was achieved using a single-mode diode-pumped solid state laser operating at $ \lambda $ = 532 nm, using a camera objective of f-number 1.8 and focal length 50 mm. The backscattered light was analyzed by a Sandercock-type (3-3)-pass tandem Fabry–Perot interferometer.

\providecommand{\noopsort}[1]{}\providecommand{\singleletter}[1]{#1}%

\section*{Acknowledgements}

This project 17FUN08 TOPS has received funding from the EMPIR programme co-financed by the Participating States and from the European Union's Horizon 2020 research and innovation programme. This project was also supported by the UK government department for Business, Energy and Industrial Strategy through NMS funding (Low Loss Electronics) and the UK national Quantum Technologies programme. A. S. and D. S. thanks EPSRC for their supports through NPIF EPSRC Doctoral studentship (EP/R512400/1) and Centre for Doctoral Training in Advanced Characterisation of Materials (EP/L015277/1) respectively. We would like to thank Pavlo Zubko for fruitful discussions on the X-ray reflectivity measurements as well as T. Dion for general discussion of the manuscript.  

\section*{Author contributions statement}

A.S. performed FMR measurements and A.S., M.C. analysed the results. S.T. and G.C. performed BLS measurements and analysed the results.  M.C. performed SQUID magnetometry measurements and D.S., M.C. analysed the results. S. Z. and X. Z. fabricated, M.S., L.H.D. and D.R. irradiated samples. E.S. performed X-ray reflectivity measurements and analysed the results. C.B. performed MFM measurements and C.B., N.S., L.W. analysed the results. H.K. and O.K. contributed to the interpretation of the results. M.C., A.S. and S.T. wrote the manuscript with inputs from the other authors. M.C. supervised the project.   

\section*{Additional information}
Springer Nature remains neutral with regard to jurisdictional claims in published maps and institutional affiliations.
\textbf{Supplementary Information}. 
\noindent\textbf{Competing interests}
The authors declare no competing interests.

\end{document}